# Robust Causality and False Attribution in Data-Driven Earth Science Discoveries

**Elizabeth Eldhose[1], Tejasvi Chauhan[1], Vikram Chandel[2], Subimal Ghosh[1,2]\*, and Auroop R. Ganguly[3,4]**

[1]Department of Civil Engineering, Indian Institute of Technology Bombay, Mumbai, India

[2]Interdisciplinary Program in Climate Studies, Indian Institute of Technology Bombay, Mumbai, India.

[3]Sustainability and Data Sciences Laboratory, Department of Civil and Environmental Engineering, Northeastern University, Boston, MA, USA

[4]Pacific Northwest National Laboratory, Richland, WA, USA

**\* Corresponding Author: Email: subimal@iitb.ac.in; subimal.ghosh@gmail.com**

## Abstract

Causal and attribution studies are essential for earth scientific discoveries and critical for informing climate, ecology, and water policies. However, the current generation of methods needs to keep pace with the complexity of scientific and stakeholder challenges and data availability combined with the adequacy of data-driven methods. Unless carefully informed by physics, they run the risk of conflating correlation with causation or getting overwhelmed by estimation inaccuracies. Given that natural experiments, controlled trials, interventions, and counterfactual examinations are often impractical, information-theoretic methods have been developed and are being continually refined in the earth sciences. Here we show that transfer entropy-based causal graphs, which have recently become popular in the earth sciences with high-profile discoveries, can be spurious even when augmented with statistical significance. We develop a subsample-based ensemble approach for robust causality analysis. Simulated data, and observations in climate and ecohydrology, suggest the robustness and consistency of this approach.

**Introduction**

The ability to establish causality or attribute an effect to a mechanism or driver has been considered the "holy grail" across disciplines ranging from physical, biological, cognitive, and ecological or environmental sciences to engineering, medicine, economics, social and management sciences, business, and even metaphysics. All of our understanding of the world initially comes through understanding the various causes and effects surrounding us. Recent developments in causal inference and their disciplinary adaptations rely on interventions under controlled conditions[1], natural experiments[2,3], and counterfactuals[4]. The 2019 Sveriges Riksbank Prize in Economic Sciences in Memory of Alfred Nobel was awarded to Abhijit Banerjee, Esther Duflo, and Michael Kremer for their work with randomized controlled trials (RCTs) in developmental economy[1], and in 2021 to Joshua D Angrist and Guido Imbens[5] for their methodological contributions to the analysis of causal relationships based on natural experiments. RCT owes its origins to developments in the late 18$^{th}$ to early 19$^{th}$ centuries, assumed its modern form through the work of Austin Bradford Hill in the mid-20$^{th}$ century[6], and has long been considered a standard approach in areas such as medicine and drug development, as well as transportation[7], criminology[8], and education[9,10]. However, in the earth sciences, where interventions and trials, and even natural experiments and counterfactuals, may be infeasible or impractical, data-driven methods such as probabilistic causation (e.g., based on Bayesian statistics)[11], causal do-calculus (e.g., causal Bayesian networks)[12,13], and structure learning (e.g., based on Granger causality (GC) and vector autoregression)[14–17] have proved useful. Mutual Information (MI) based measures have been proposed to extract generalized dependence patterns; however, persistent challenges remain around the equitability of estimation and the stability of estimates, especially with finite and noisy data[18,19]. The Schreiber transfer entropy[20], a Kullbach-Leibler distance of transition probabilities[21], can be shown to be equivalent to conditional mutual information (CMI). While the MI is a standard information-theoretic measure of generalized dependence, the CMI or transfer entropy (TE) is a measure of conditional dependence. The CMI with time delays can be shown to be a nonlinear extension to GC[22,23] and has been used in areas such as neurophysiology[24], biomedical engineering[25], climate science[26], laser dynamics[27], and finance[28].

The climate and earth science literature has followed the developments in the field of causal analysis with certain important caveats. First, given the complex dynamical nature of the climate, ecological, and water systems, natural experiments, interventions, and controlled trials are often impractical and, at times, unethical even if feasible. The extant literature has characterized how human interventions through land-use change[29] and carbon emissions[30] have already affected the earth's climate by contributing to irreversible observed changes in the system. The alternatives are data-driven approaches[31] and numerical simulation experiments[32,33]. A ground-breaking work by 2021 Physics Nobel Prize winner Klaus Hasselmann on climate attribution has been directional to the research community. Attribution studies in climate use variants of Hasselmann's "optimal fingerprinting" method[34,35], which involves a linear regression of historical climate observations on corresponding output from numerical climate models, with increasing sophistication and testing over the last half-century. However, knowledge gaps preclude developing numerical models in many earth science problems. Furthermore, numerical experiments are computationally time-consuming. While insights into system behavior can also be drawn through direct analysis from model equations[36], they may be limited by the assumptions and knowledge gaps inherent in model-building. Meanwhile, with the growing availability – even deluge – of data for the land, ocean, and atmospheric variables from field-based in-situ instrumented records, remotely sensing observations such as from satellites, and archived model simulations following data assimilation, the data-driven causal discovery has been gaining popularity. However, even as data-driven causal inference is being seen as critical to predictive understanding in climate and earth science and for risk-informed policy and decision making, the need for examining and reducing the potential for spurious causality detection and false discoveries of cause and effect is becoming crucial. Here we examine the growing use of information-theoretic approaches for causal analysis in the earth sciences in this context.

**State-of-the-Art and Key Gaps**

Current Detection and Attribution (D&A) in climate typically use numerical earth systems model simulations to estimate the fingerprints of climate change[34,35]. Climate models have been shown to be credible tools to study how the earth system responds to a controlled perturbation,[37,38] but their performances are often limited in capturing several stakeholder-relevant processes and feedbacks[39]. In earth system science, studying the interactions within and among the subsystems remains challenging as the processes may exhibit significant spatial variability and heterogeneity from local (e.g., soil-plant-atmosphere interactions) to global (e.g., climatic teleconnections) scales and temporal variability from daily (e.g., temperature-extreme precipitation association) to multi-decadal scales (e.g., Atlantic multi-decadal oscillations and its global impacts). Prior investigations of these interactions have primarily relied on correlation-based analyses[40–45]. However, as has been pointed out in the data and climate science literature, conflating correlation with causation may lead to serious flaws in our understanding. The other criticism of correlation-based studies is that they can only detect linear associations between the variables and hence can be misleading in the earth sciences, where nonlinearity is nearly ubiquitous. Thus, Mutual Information (MI), a nonlinear extension of correlation as explained in the previous section, has been extensively used in earth sciences[46,47]. The nonlinear dynamical literature has occasionally attempted to improve MI significantly. Thus, the Maximal Information Coefficient (MIC)[18] was proposed as an extension of MI. However, the superiority of MIC over MI has been contested in the literature[19], and estimation processes may not be stable with finite data and noise in the presence of autocorrelations, which is often the case in climate and earth sciences. On another note, directional associations inferred from linear correlation[40–43,45] or the generalized nonlinear dependence estimations based on the MI do not imply causation. Based on these considerations, data-driven causal discovery-based approaches[17,48–52] have become increasingly popular in the earth sciences leading to high-impact claims pertaining to causal inferences and earth science discoveries. However, a detailed examination of the robustness of the methods and hence the credibility of the claims has been lacking.

Causality analyses can be used to estimate the general functional relationship between system variables[53]. GC (GC)[54] (see Data and Methods) is a widely applied data-driven approach to detect causality in earth science[14–16,48,55]. A time-series variable X is said to Granger cause another variable Y if the past values of X could improve the prediction of Y conditional to its own past. GC is thus a prediction-based technique implemented using a linear autoregressive model, which is not expected to fully capture the inherent complexities in earth systems, especially when there is non-separability, weak coupling, and nonlinearity. Transfer entropy (TE),[20] (an information-theoretic approach discussed previously, see Data and Methods for details) is a nonlinear variant of the traditional GC approach[22,23], defined as the information flow from source variable to target conditioned on the past information of the target. Since TE deals with the probability density functions of time-series data, it provides a unifying way to deal with variables of different units and scales (e.g., precipitation with units of volume per unit time and water storage with units of volume)[56,57]. Causal detection using TE remains an emerging topic in earth and climate sciences[58,59] although it has been widely applied in other disciplines such as to study financial time series[28], chemical processes[60], and neuroscience[61]. The outstanding gap, however, is that the robustness and stability of TE estimations and TE-based causal graphs have not been adequately studied, given the expected challenges in estimation and detection processes in earth systems. Detection of spurious causalities with TE-based approaches is a possibility[62], especially for time series where the data availability period is limited. Satellite observations and related data products, such as MODIS, AMSR-E, and GRACE, are available during the post-2000 period with a temporal resolution of a month, thus resulting in a sample size of approximately 250, which are further characterized by higher noise-to-signal ratios. Therefore, based on the information-theoretic literature on TE and TE-based causal graphs, the possibility that TE-based insights in the recent climate and earth science literature may be misleading, owing to spurious TE-graph links[58,63–66], needs to be examined. However, testing the robustness of TE-based inferences and discoveries, while critical, is not straightforward owing to incomplete process knowledge and a relative shortage of time series data.

**Ensemble Approaches for Robust Estimation**

Ensemble approaches have been empirically shown to improve the robustness of model performance across disciplines[67]. Here, we propose an approach similar to bootstrapping to identify robust (TE-based) causal links in a multivariate system (see Data and Methods: *Ensemble Approach for Robust Estimation of Transfer Entropy* and Supplementary Figures 1 and 2). The approach checks the consistency of statistically significant links across continuous sub-samples generated from a sample. To validate the proposed method, we first synthetically constructed multivariate time-series data that are designed to be independent, linearly correlated, or with nonlinear dependence. We then used the bivariate TE approach to delineate the causality between variables in these systems and found that spurious links exist even after using a large sample and applying a statistical significance test. We could eliminate such links by checking consistency across continuous sub-samples generated from the sample. We demonstrated our approach with TE as well as GC.

Interestingly, the approach also addresses, to a certain extent, the issue of transitivity (if X causes Y and Y causes Z, then it implies X causes Z, irrespective of whether they are directly linked) and common driver (if X causes both Y and Z at different time lags, then Y and Z seems to be causally connected). We examined our approach on two natural (earth) systems: the Tropical Pacific Walker Circulation and a local ecohydrological problem. Our approach mainly focuses on eliminating false links that may be spuriously detected because of data limitations. The approach can reliably estimate causal links in earth sciences and may generalize to other disciplines.

**Experimental Simulation Design**

Our proposed approach is based on the following concept: applying a causal discovery method to multiple time-continuous subsamples of sufficient length and checking whether consistent links in graphs thus produced increase the robustness of the links estimated. We demonstrate this approach by applying TE to a hundred random sub-samples with the temporal structure preserved (200 data points out of 1000). We consider a link as robust if it appears in more than 90 percent of subsamples. In real-world problems, generating 100 subsamples might not be possible as the data length is limited. In such a case, we show that this method is robust with a shorter number of subsamples from the data.

We generated three multivariate synthetic systems with already known relationships among their components to test the robustness check. Further, we test the robustness check on two real-world examples already established in literature[51,56,68].

First, we generated a random system (*system A*) consisting of four independent and normally distributed variables, each with a zero mean, unit standard deviation, and a sample size of 1000. The variables are denoted by *X(t), Y(t), Z(t),* and *W(t)*. The second generated system (*system B*) has the same variables *X(t), Y(t), Z(t)*, and *W(t)* with dependencies among themselves (Eq. 1-4). A sample of 1000 data points was generated using these equations, and the first 100 values were discarded to obtain stable series.

$$X(t) = 0.4\, Z(t-1) + \eta_t^x \tag{1}$$

$$Y(t) = 0.6\, X(t-3) + 0.09\, W(t-2) + \eta_t^y \tag{2}$$

$$Z(t) = 0.7\, Y(t-2) + \eta_t^z \tag{3}$$

$$W(t) = 0.5\, X(t-1) + \eta_t^w \tag{4}$$

Here, $\eta_t$ is a random normal variate with mean 0 and standard deviation 1. We further introduced nonlinearity to the system by replacing Eq. (1) with Eq. (5). This system (Eq. 2-5) is referred to as *system C*.

$$X(t) = 0.4\, Z(t-1)^2 + \eta_t^x \tag{5}$$

We develop causal graphs for the three systems mentioned above using pairwise TE, grounded on information theory. The equations (Eq. (1)-(5)) defining the systems provide the actual links, whereas the TE-based graphs comprise of statistically significant links (methods). We applied the proposed robustness check on above mentioned three systems and validated our results using links from equations.

**Simulation Results and Discussion of Lessons Learned**

We first computed the TE up to four lags between all possible pairs of *system A* with a sample size of 1000 (Figure 1(a), explained in Experimental Design), and the causal graph for the system is presented in Fig 1(b). We evaluated TE using the fixed binning approach (methods), and we showed that links remain stable around the selected number of bins (±2 bins) (Supplementary Figure 3). Ideally, there should not be any link among the independent variables. Still, spurious links appear at both higher (lag 4; Fig 1(f)) and lower lags (lag 1 and 2; Fig 1(c) and (d)), even after testing for statistical significance (methods). We generated graphs for 100 sub-samples, each of size 200, from system A (Experimental design). The fraction of times statistically significant links are generated using TE is shown in Fig 1(g)-(j) for lag 1 to 4, respectively. After sub-sampling, we find that all the spurious links appear in less than 40% of subsamples. If we consider the links appearing in more than 90% of the subsamples as robust links, the method does not produce any links for any of the variable pairs (Fig 1(k)-(n) for lag 1 to 4 respectively), which is the desired outcome for *system A*.

To understand if such a robustness test applies only to the TE or can be considered a generalized approach for other causality detection methods, we reperformed the analysis using GC on *system A* (Supplementary Figure 4). The GC applied to *system A* also generates a few spurious links (Supplementary Figure 4(b)-(f); all at lag 2), which are eliminated after the robustness check (Supplementary Figure 4(k)-(n)).

We applied a similar approach to *system B* (Supplementary Figure 5). The TE-based graph reveals a well-connected network with multiple spurious links (Supplementary Figure 5(b)) when applied to the sample of size 900. The spurious links are generated more in the higher lags (lag 2, 3, and 4, Supplementary Figure 5(d)-(f)). A higher number of links at higher lags could be indirect links; for example, the link at lag 4 $Z \to Y$ is due to link $Z \to X$ at lag 1 and $X \to Y$ at lag 3. The link $W \to Z$ at lag 4 is because of the combinations of two lag 2 links, $W \to Y$ and $Y \to Z$. However, the link $Z \to W$ at lag 4 is fully spurious and cannot be explained. We applied our proposed robustness check, and the results are presented in Supplementary Figure 5 (g)-(n). Our approach could eliminate all the spurious links, including the indirect links (Supplementary Figure 5 (k)-(n)). However, the method failed to detect an actual link $W \to Y$ at lag 2 (Supplementary Figure 5(l)), which is a weaker link by nature. We found that the link $W \to Y$ appeared in more subsamples (more than 50%) compared to other spurious links (Supplementary Figure 5(h)). However, considering a higher threshold of 90% eliminates the link at lag 2 from the final graph.

We applied the bivariate GC to *system B* and found a well-connected graph derived from GC. There are spurious links at higher lags (Supplementary Figure 6) because we use an autoregressive model. There are also links representing indirect connections, as in the case of TE. The robustness check applied to the sub-samples could eliminate several indirect links such as $Z \to W$ in lag 2 (Supplementary Figure 6(l)), $Y \to X$ in lag 3 (Supplementary Figure 6(m)), $W \to Z$ in lag 4 (Supplementary Figure 6(n)).

TE is designed to derive nonlinear causal associations; hence, we apply our approach to *system C* (Figure 2). The TE applied to a sample of size 900 reveals a well-connected graph (Figure 2 (b)) similar to *system B*. For this case, all the spurious links, such as $Z \to W$ at lag 2, $Y \to X$ at lag 3, $Y \to W$, and $W \to Z$ at lag 4 are indirect links or transitive links that occur through other variables (Fig 2c, 2d, 2e, 2f). The robustness check applied to smaller subsamples eliminates a couple of indirect links at lag 4 (Figure 2 (n)). However, some indirect links remain in the graph since we use bivariate analysis. Similar results were obtained when the robustness check was applied to the nonlinear system with the GC approach (Supplementary Figure 7).

The TE worked well for the nonlinear system with a high sample size (900). To test our approach on a limited sample size, we picked up a sample of size 200 from the nonlinear system (*system C*) and applied TE to derive the graph. The results are presented in Figure 3. When TE is applied to the sample of size 200, spurious links start appearing from lag 1 (Figure 3 (c)), $X \to Z$. The number of spurious links is very high at higher lags of 3 and 4, such as $Y \to Z$, $Z \to X, X \to W, W \to X$ all at lag 3. To do a consistency check for testing robustness, we took three sub-samples of 100 data points: the first 100, the middle 100, and the last 100, with overlaps of 50 between them. Figures 3(g)-(j) show the number of samples where each link appeared for lags 1 to 4, respectively. For a 90% consistency across subsamples like in previous examples, a robust link should be present in all the sub-samples. The robust links are shown in Figures 3 (k)-(n) for lags 1 to 4, respectively. The consistency check for robustness eliminated all the spurious links. The method could identify all the direct links except for $W \to Y$ at lag 2. All the indirect links were eliminated except for $Y \to X$ at lag 3, originating from $Y \to Z$ at lag 2 and $Z \to X$ at lag 1.

After applying the robustness check approach to synthetic systems, we applied our approach to real-world earth system data.–First, we tested our approach on the Pacific Walker Circulation[51,68]. We apply the robustness check on the TE-based causal graph between monthly pressure anomalies over the western pacific region ($WPAC$) and monthly surface air temperature over the central pacific ($CPAC$), east pacific ($EPAC$), and Atlantic ($ATL$) region. The connections between the variables obtained by an advanced causality approach, Peter Clark Momentary conditional independence (PCMCI), by Runge et al.[51] are shown in Fig. 4(b). When TE was applied, we found an almost fully connected network (Fig. 4(c)). As discussed in Runge et al.[51], following the scientific analysis of Alexander et.al[69], many of the links are spurious and were probably generated because of high correlation (shown in Runge et al.[51] Figure 4A of the reference). Ideally, Pacific Walker Circulation or the Pacific overturning cell arises due to the temperature and pressure relationships between the east and west Pacific. The moist air from the east (EPAC) is carried to the west (to WPAC through CPAC) through the tropics, near the surface. There it dries and is carried upwards. The loop is completed by this air sinking eastwards (WPAC to CPAC and EPAC). An atmospheric bridge from the Pacific to the Atlantic SST (ATL) also exists. The method identifies links from CPAC and EPAC to ATL, showing the atmospheric bridge. The reverse of the atmospheric bridge is not true [69]. Hence, there should not be any outgoing links generated from ATL. The application of TE shows the links from ATL consistently for all the lags 1 to 4 (Figure 4(d)-(g)). After applying the robustness check with 100 subsamples (fraction of cases when links appeared shown in Fig. 4(h)-(k)), we found that all the links generated from ATL disappeared (Fig. 4(l)-(o)). The method correctly identifies the links from $EPAC \to CPAC$, $CPAC \to WPAC$ and indirect link $EPAC \to WPAC$. The eastward sinking is also represented by $WPAC \to CPAC$ and $WPAC \to EPAC$. We also found links from CPAC to EPAC. This link could be generated because of El-Nino and Modoki events when SST anomalies are reversed. A detailed explanation of the same requires separate investigations of El-Nino, La-Nina, and Normal years, which is out of scope for the present work.

After examining the proposed approach for a large-scale climate phenomenon, we applied it to a local scale eco-hydrological process used by Ruddell and Kumar[56]. We considered the half-hourly data for five ecohydrological variables; precipitation ($P$), surface layer soil water content ($\theta$), sensible heat flux ($\gamma_H$), latent heat flux ($\gamma_{LE}$), and surface layer soil temperature ($\theta_S$), for July 2003 (1460 data points). Though the sample size looks large, the data is half-hourly; hence, considering 10 days needs 480 data points. Ruddell and Kumar suggested a minimum of 500 data points to apply TE for generating the causal graph, termed process network. Applying a pairwise TE to the entire sample 4 lags shows an almost fully connected network (Figure 5(a)). The network consists of land-atmosphere feedback links from $\theta$, $\theta_S$, and $\gamma_H$ to $P$. Land-atmospheric feedback from a point scale value that too within 2 hours (4 lags) to $P$ is nearly impossible. These feedback links were present in all the lags (Fig. 5(b)-(e)). For the application of the robustness check, we could generate three subsamples, each of size 500. We found that the land-atmosphere feedback links are not consistent at lags 1 to 4 (Fig. 5(f)-(i)). After applying the robustness check, the graphs generated at different lags show strong connections among land variables as expected but no feedback links. This analysis further establishes the usefulness of the proposed approach to test the robustness of causal links derived statistically.

We do not expect any statistically significant links when applied to the random system driven by noise. However, results show spurious links appearing at various lags. We performed a consistency check on 100 subsamples, and spurious links disappeared, showing that our proposed robustness check was able to eliminate them. For the linear and nonlinear systems, we expect TE to detect indirect and confounding links (as TE used here is bivariate) but no spurious links. The results from conventional TE analysis showed a few spurious and many indirect links. Our method eliminated some indirect links and all spurious links.

While applying TE to the Pacific Walker Circulation data and the local ecohydrological problem, we expected TE to capture all the actual links, some spurious links because of observational errors, and indirect links. Instead, we found a fully connected network for both systems except for one link. Our robustness check was able to eliminate all the spurious links. Our approach also delineated the atmospheric bridge between the Central Pacific and the Atlantic[69]. In the local ecohydrological system, we got feedback links to precipitation from temperature, soil moisture etc., at sub-hourly to hourly time lags (30, 60, 90, and 120 minutes), which are most likely spurious and were removed once the consistency check was done.

To summarise, recent literature shows a significant increase in statistical analyses in earth science; however, most do not ensure robustness and consistency. There is a growing importance of causal analysis in earth system science, specifically in climate science, to understand processes that are not fully understood. They are also equally important for developing novel detection and attribution techniques. A link generated spuriously because of a poor sample or a non-robust statistical approach may lead to wrong scientific knowledge about a process. Here, we developed a technique of consistency check across sub-samples to test the robustness. After successfully applying the technique to synthetic multivariate systems, we tested the same for large-scale and small-scale, well-established earth system processes. We found the approach worked well. We recommend such a test to be applied for the robustness check of causal links in the growing number of statistical causal analyses in different disciplines, including earth system science.

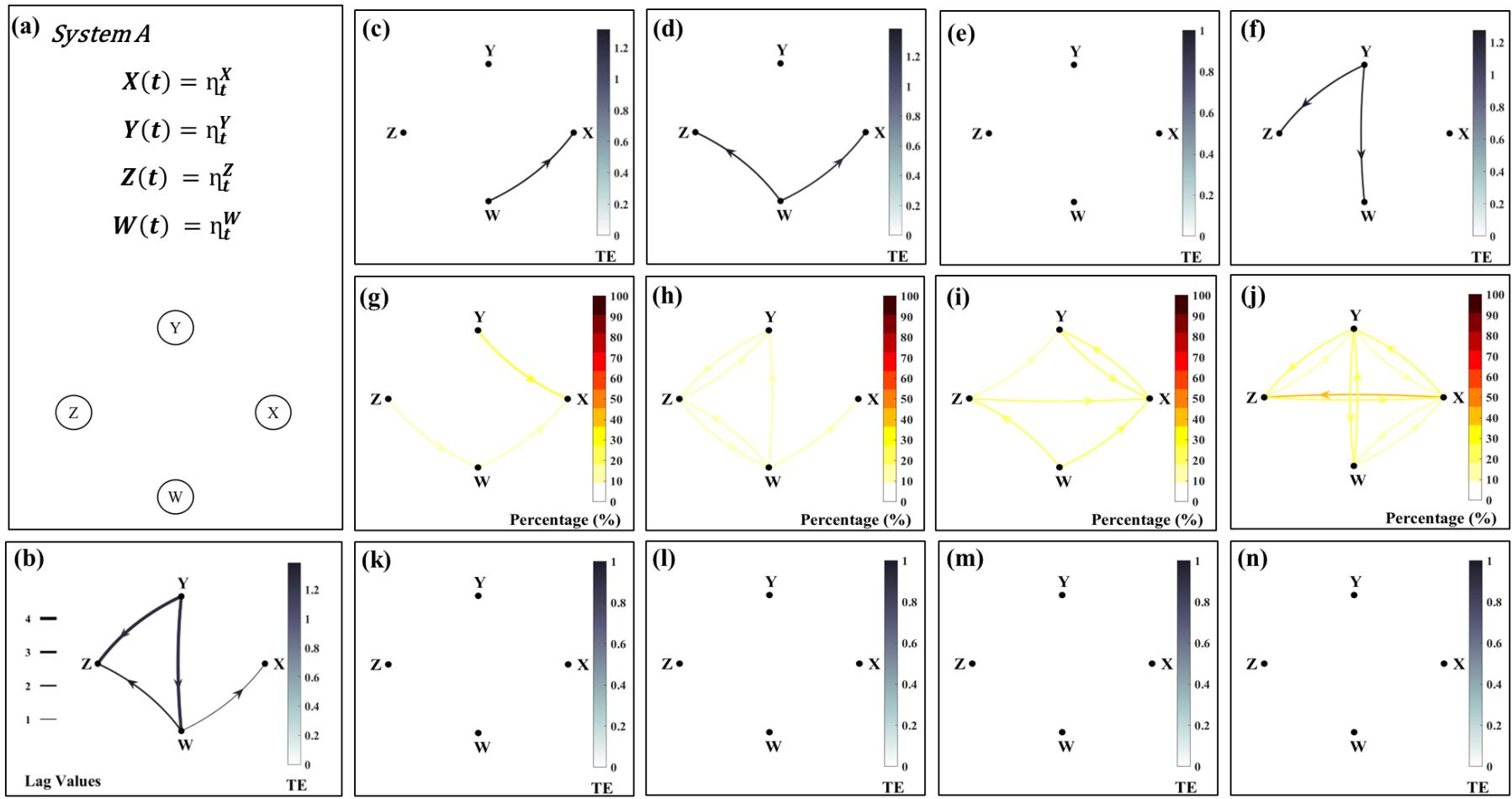

*Figure 1:* (a) System A of four independent variables; X, Y, Z, and W. Each variable is a time series of normally distributed random numbers with 1000 data points. Plot (b) shows a directed graph with the significant transfer entropy links up to lag 4, obtained from the sample of size 1000. Lag-wise transfer entropy links are presented in (c) Lag 1, (d) Lag 2 (e) Lag 3, (f) Lag 4. Figures (g) -(j) denote each link's appearance percentage at the four lags after applying transfer entropy to 100 subsamples of size 200 each. Correspondingly, the links that appeared above 90% times at four lags are shown in (k) - (n)

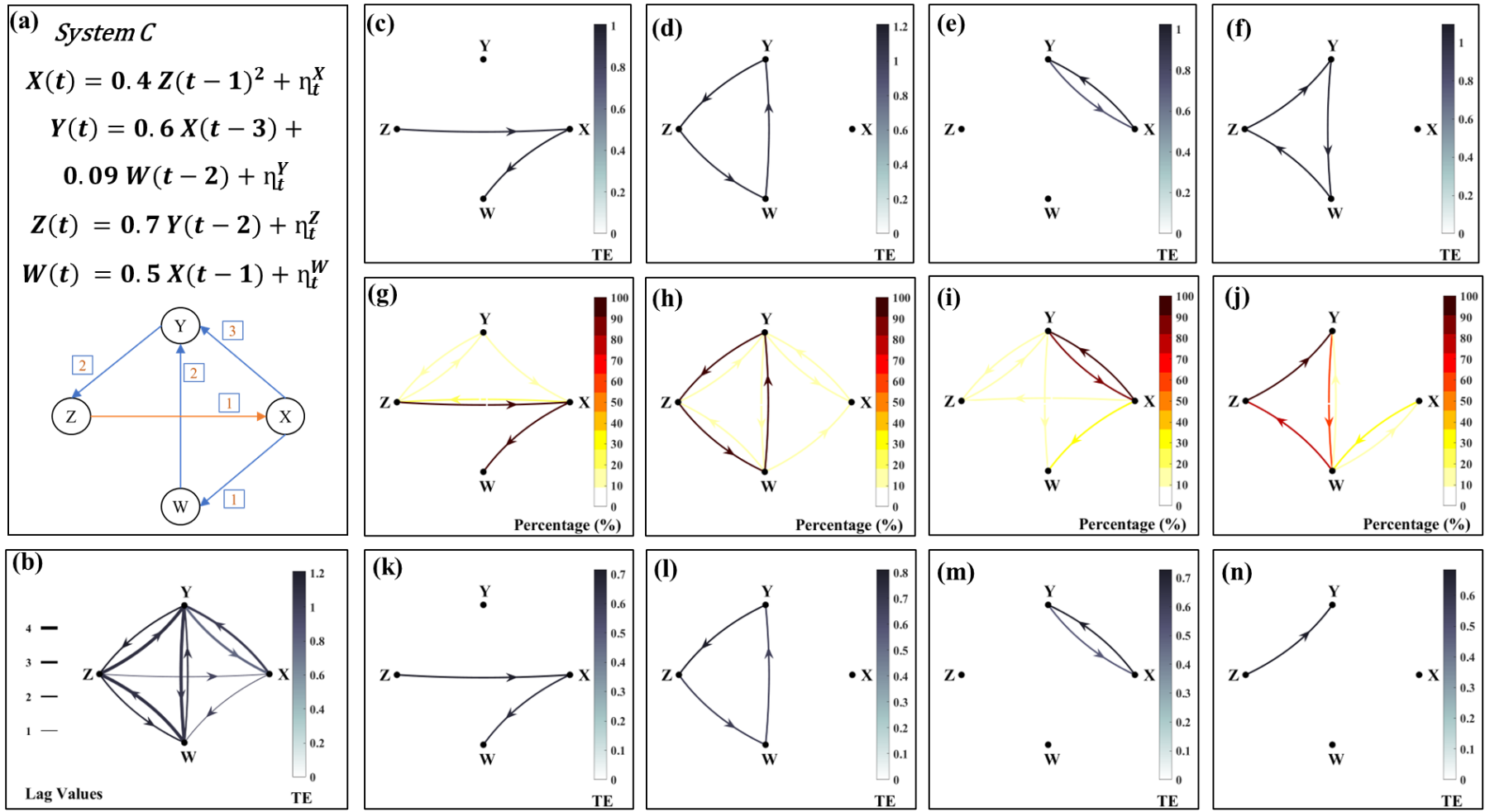

*Figure 2:* (a) System C of four variables, X, Y, Z, and W, of 900 data points each. A sample of size 1000 is generated with the equations mentioned in (a). After discarding the first 100 values, the rest 900 values are used for further analysis. Plot (b) shows a directed graph with the significant transfer entropy links up to lag 4, obtained from the sample of size 900. Lag-wise transfer entropy links are presented in (c) Lag 1, (d) Lag 2 (e) Lag 3, (f) Lag 4. Figures (g) -(j) denote each link's appearance percentage at the four lags after applying transfer entropy to 100 subsamples of size 200 each. Correspondingly, the links that appeared above 90% times at four lags are shown in (k) - (n)

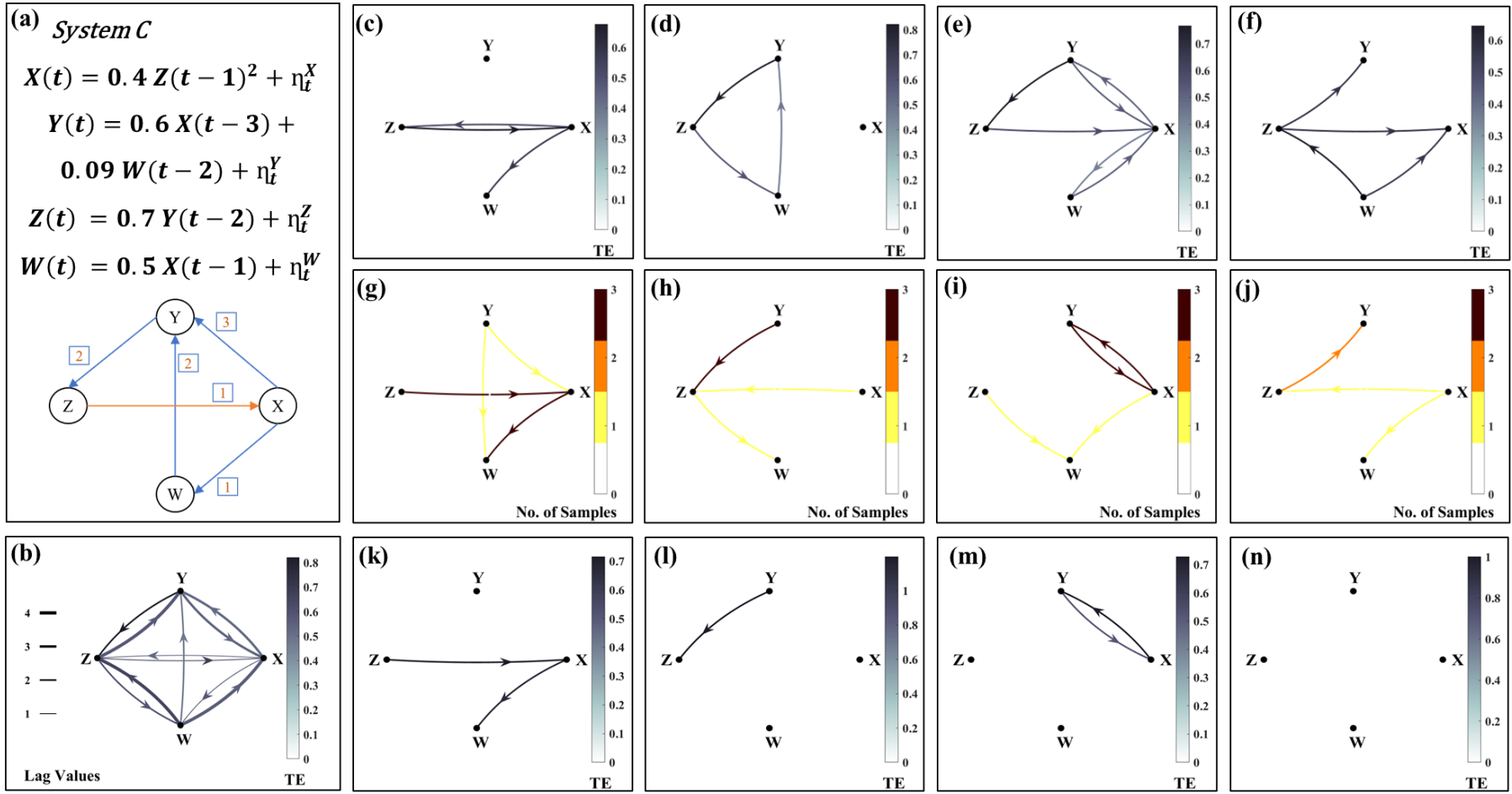

***Figure 3:*** *A random subsample of 200 data points is chosen from System C (Figure 2a). Plot (b) shows a directed graph with the significant transfer entropy links up to lag 4, as obtained from the sub-sample of size 200. Lag-wise transfer entropy links are presented in (c) Lag 1, (d) Lag 2 (e) Lag 3, (f) Lag 4. We further generate three subsamples, each of size 100 from the sample of size 200. Figures (g) –(j) denote the no of subsamples in which individual links appeared. Correspondingly, the links that appeared in all the three samples at the four lags are shown in (k) - (n)*

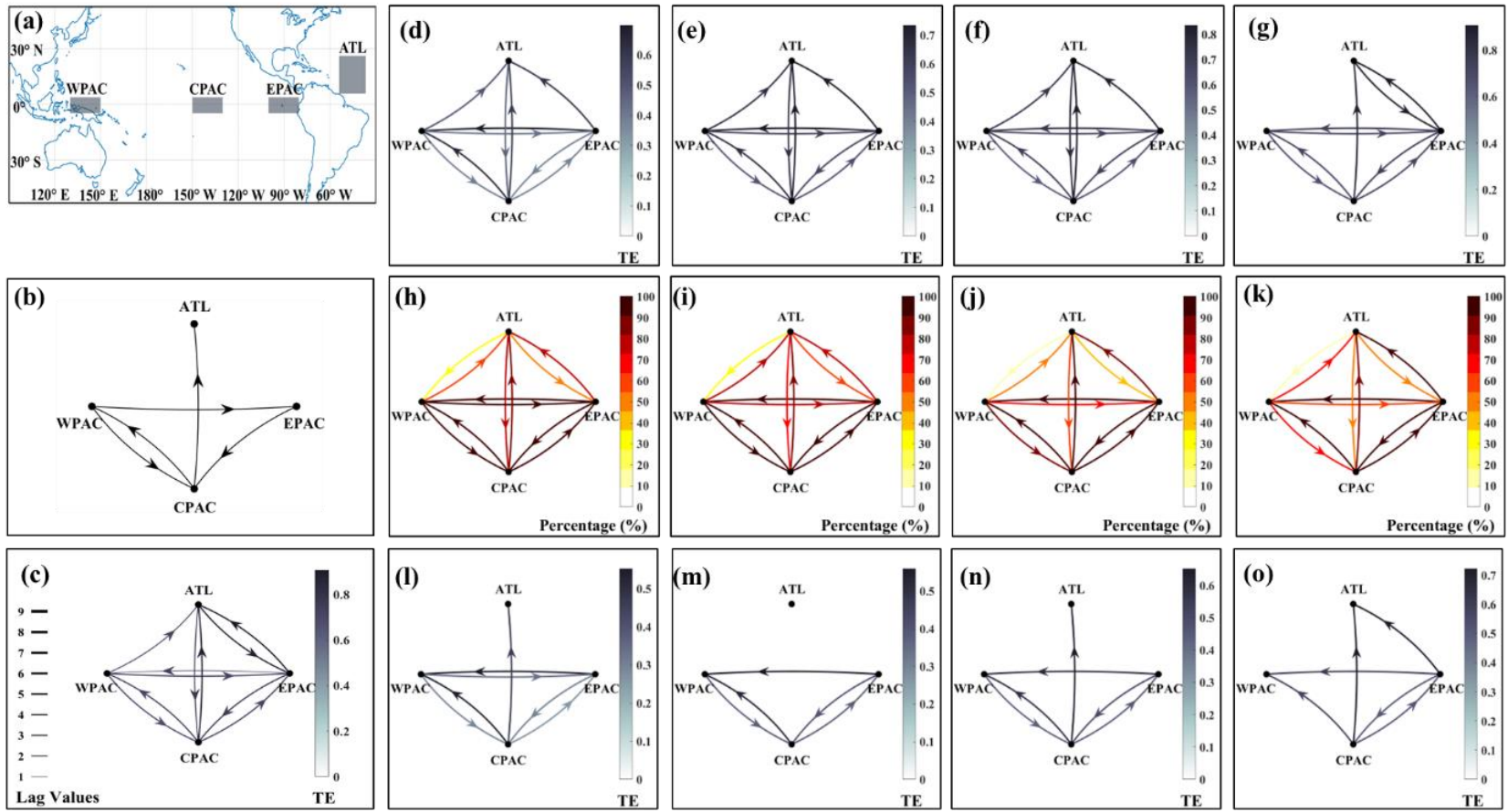

*Figure 4*: Tropical Pacific Walker Circulation for the period 1948–2012 (T = 780 months) used as an example. (a) the grey shaded boxes show the regions used for analysis. Variables considered: monthly surface pressure anomalies in the West Pacific (WPAC) and surface air temperature anomalies in the Central (CPAC) and East Pacific (EPAC), and tropical Atlantic (ATL). (b) Tropical Pacific Walker Circulation as explained in Runge et al. {21}. (c) A directed graph showing the significant transfer entropy links up to lag 4. Lagwise transfer entropy (d) Lag 1, (e) Lag 2 (f) Lag 3, (g) Lag 4. Figures (h) -(k) denote each link's appearance percentage at the four lags after analyzing 100 subsamples. Correspondingly, the links that appeared above 90% times at the four lags are shown in (l) - (o)

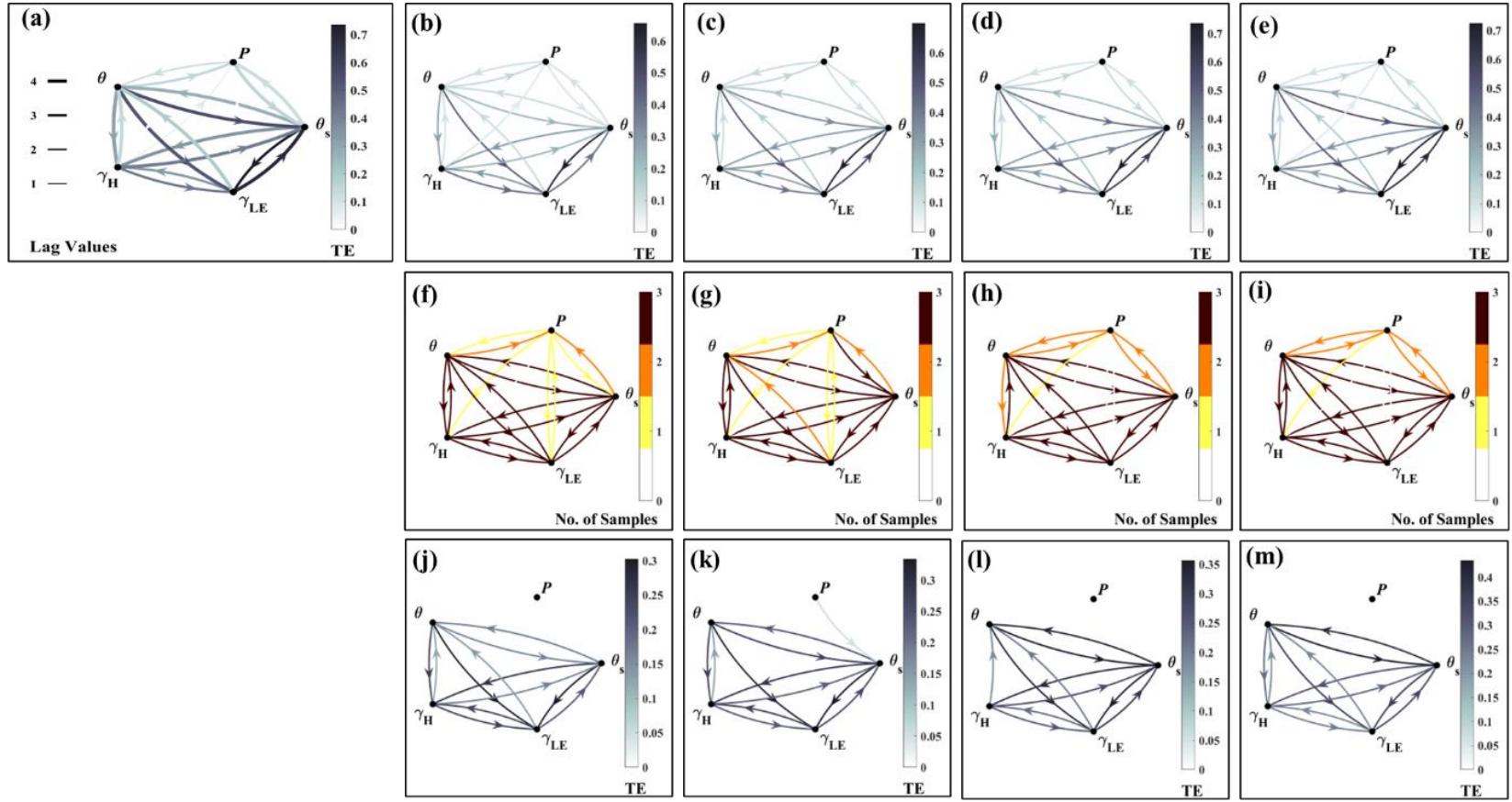

*Figure 5:* A local ecohydrological system at Bondville, USA, for July 2003 FLUXNET data (half-hourly data) used as an example. Variables considered are Precipitation (P), surface layer soil water content (θ), sensible heat flux ($\gamma_H$), latent heat flux ($\gamma_{LE}$), and surface layer soil temperature ($\theta_s$). (a) A directed graph showing the significant transfer entropy links up to lag 4 considering all the 1460 data points. Lagwise transfer entropy (b) Lag 1, (c) Lag 2 (d) Lag 3, (e) Lag 4. We further generate three subsamples, each of size 500, from the data. Figures (f) -(i) denote the no of subsamples in which individual links appeared. Correspondingly, the links that appeared in all the three subsamples at the four lags are shown in (j) - (m)

## Data and Methods

### Data

We generated data for the three synthetic systems with predefined relationships. To test our method on Pacific Walker Circulation, we collected the climate data from the National Centers for Environmental Prediction–National Center for Atmospheric Research (NCEP–NCAR) reanalysis [70] dataset for the period 1948-2012 (780 months). The regions used for analysis are shown in Fig 4(a). We pre-processed the data by removing the linear trend and subtracting the mean seasonal cycle. For the local ecohydrological system, we used the data from Ruddell and Kumar[56].

### Methods

### Transfer Entropy

We used the information-theoretic approach, transfer entropy (TE) [20], to delineate the causal links between variables. It is a Shannon entropy-based statistic. Let $X$ denote a discrete time series variable with n data points such that $X = \{x_1, x_2, ..., x_n\}$ with its marginal pdf $p(X)$, then Shannon's entropy (a measure of uncertainty associated with a variable) is given by,

$$H(X) = -\sum_{i}^{m} p(x_i) \, log\,[p(x_i)] \tag{6}$$

Mutual Information (MI; Information redundancy) is the reduction in this uncertainty and can be defined as the predictive information that the variables hold for each other. It is a symmetric quantity and can be given by,

$$MI(X_t; Y_t) = H(X_t) + H(Y_t) - H(X_t, Y_t) \tag{7}$$

, where $H(X_t, Y_t)$ is the joint variability of both the variables or the joint information entropy.

$$H(X_t, Y_t) = \sum_{i,j=1}^{m} p(x_i, y_j) \, log\,[p(x_i, y_j)] \tag{8}$$

TE is the conditional mutual information that gives the information flow from source variable to target variable conditioned on the target's past information. Information flow refers to the predictive knowledge of one variable contained in another.

$$T_E(X_{t-\tau_x}; Y_t) = I(X_{t-\tau_x}; Y_t \mid Y_{t-\tau_y}) \tag{9}$$

TE gives the reduction in uncertainty of the current state of $Y_t$ with the knowledge of $X_{t-\tau_x}$ (variable $X$ considered at a lag $\tau_x$) over the knowledge of $Y$'s past ($Y_{t-\tau_y}$; $Y$ at lag $\tau_y$). Here for ease of computation, we considered $\tau_x$ and $\tau_y$ to be same. Since TE detects the predictability of a variable by another over its past, it could be said equivalent to GC with the advantage of detecting nonlinear dependencies[22,23]. We computed TE from the component Shannon entropies using Eq. 10

$$T_E(X_{t-\tau}; Y_t) = -H(Y_{t-\tau}) + H(X_{t-\tau}, Y_{t-\tau}) + H(Y_{t-\tau}, Y_t) - H(X_{t-\tau}, Y_{t-\tau}, Y_t) \tag{10}$$

TE ranges from 0 to min $(H(X_t), H(Y_t))$. Here we used the fixed binning technique[56,71] to estimate the marginal and joint probability distributions owing to its simplicity, and the number of bins was decided using Scott's rule and implemented in MATLAB. For a time-series with $l$ elements and standard deviation $\sigma$, the bin width using Scott's rule is given by,

$$bin\ width = \frac{3.5\ \sigma}{\sqrt[3]{l}} \qquad (11)$$

The minimum bin number obtained among all the variables in the system is finalized for estimating the marginal and joint probabilities. We also performed a sensitivity analysis on the number of bins by considering $\pm 2$ bins to see if the choice of bins affects the detection of links using TE. Supplementary Figure 3 shows the TE graph for *system C* (Eq. (2)-(5)) with bins in and around the bin number obtained using Scott's rule. The figure shows that the changes in the bin number used for fixed binning do not cause many changes in the links detected.

The method of random shuffled surrogates is used to ensure the statistical significance of the links. The time series variable is shuffled, destroying the time relationships, and MI is calculated for 100 such realizations. The mean and standard deviation of the computations are calculated, and the standard t-test is carried out at 95% confidence level. This is supposed to eliminate the couplings that occur by random chance between two unrelated time series. TE is computed only if the MI between the variables is significant, as TE is conditional mutual information (CMI).

**Granger Causality**

Granger Causality (GC) provides a framework for identifying causation between time series variables using the idea of predictability. If a variable X's past could improve the prediction of another variable Y's present than the prediction with its own past, then variable X is said to granger-cause Y. It assesses how the past of "cause" variables affects the predictive distribution of "effect" variables.

The test is based on the linear regression model:

$$y_i = \alpha_0 + \sum_{j=1}^{m} \alpha_j y_{i-j} + \sum_{j=1}^{m} \beta_j x_{i-j} + \varepsilon_i \qquad (12)$$

Here, the $\alpha_j$ and $\beta_j$ are the regression coefficients and $\varepsilon_i$ is the error term. The test is based on the null hypothesis: $H_0 : \beta_1 = \beta_2 = \cdots \beta_m = 0$. X is said to Granger-causes Y when the null hypothesis is rejected. An F-test is used to determine whether there is a significant difference between the full model and the reduced model (where all $\beta_j = 0$).

**Ensemble Approach for Robust Estimation of Transfer Entropy**

Ensembles are well-established methods to perform better, both statistically and computationally[72,73]. Ensemble estimation was previously used to improve the accuracy of kernel density-based plug-in estimators of MI[74]. Here, we are interested in the robustness of a causal link estimated in a multivariate system, proposing an approach similar to bootstrap aggregating or bagging. The approach checks the consistency of statistically significant links (methods) across random continuous sub-samples generated from a dataset.

For a dataset $D$ of length $l$ consisting of two time-series variables, let $TE_{[p]}(D(X) \rightarrow D(Y))$ denotes the TE from source variable $X$ to target variable $Y$ at lag p.

$$D = \{(X_i, Y_i)_{i = 1, 2, ..., l}\} \tag{13}$$

For bagging, $n$ random continuous subsamples $((d_j)_{j=1,2,...,n})$, each with a length $q$, are generated from the original dataset $D$.

$$d_j = \{(X_k, Y_k)_{k = r, r+1, r+2, ..., r+q}\} \tag{14}$$

where $m < l$ and $1 \leq r \leq (l - q)$, and $r$ is chosen randomly. Now, we find whether link $X \rightarrow Y$ (denoted by $L^{X \rightarrow Y}$) exists in each of these samples.

$$L_j^{X \rightarrow Y} = \begin{cases} 1, & \text{if } TE_{[p]}(d_j(X) \rightarrow d_j(Y)) \text{ is statistically significant} \\ 0, & \text{otherwise} \end{cases} \tag{15}$$

We consider the link $X \rightarrow Y$ as robust only if it is statistically significant in more than 90% of the subsamples. Thus, the ensemble $TE_E$ can be represented as:

$$TE_E = \begin{cases} 1, & \text{if } \sum_{j=1}^{n} L_j^{X \rightarrow Y} \geq 0.9 \times n \\ 0, & \text{otherwise} \end{cases} \tag{16}$$

The final ensemble $TE_E$ gives a consistent measure of the link present at a particular lag $p$, throughout the sample. A flow chart for the same is shown in Supplementary Figure 1.

We show the performance of an ensemble estimator in a bivariate system with the help of simulations. We generated 10000 samples using equations 17 (linear) and 18 (nonlinear) for data lengths 100 and 1000 with varying $(m/\epsilon)$ ratios.

$$Y(t) = m \times X(t - 1) + \epsilon \times \eta(t) \tag{17}$$

$$Y(t) = m \times X(t - 1)^2 + \epsilon \times \eta(t) \tag{18}$$

, where $m$ is the signal coefficient, $\epsilon$ is the coefficient of noise term, and t is the timestep. Ideally, at lag 1, TE after statistical significance ($TE_{[1]}$) should be detecting links from X to Y for all the samples generated; if not detected, they are false negatives (type II errors). For $TE_{[1]}$ the false negatives (FN) and true positives (TP) would sum up to the total number of samples used (N = 10000). Similarly, at lag 2, a link detected in any sample from X to Y would be a false positive (FN; type I error), and the rest are true negatives (TN). For data lengths 100 and 1000, we computed the false negative and false positive rates using equations 19 and 20, considering the 10000 samples generated.

$$False\ Negative\ Rate\ (FNR) = \frac{FN}{FN + TP} = \frac{\sum_i^N TE_{[1]}(X_i \to Y_i) = 0}{N} \quad (19)$$

$$False\ Positive\ Rate\ (FPR) = \frac{FP}{FP + TN} = \frac{\sum_i^N TE_{[2]}(X_i \to Y_i) = 1}{N} \quad (20)$$

Supplementary Figure 2 shows the error rates as a function of varying $(m/\epsilon)$ ratios for both linear (Fig 2(a), 2(b)) and nonlinear systems (Fig 2(c), 2(d)). The error rate (both false positive and false negative) is more for data length 100 (blue lines), which shows the explicit dependence of TE on data lengths. When the data length is reduced, the probability of getting a spurious link increases.

Let us say we have a parent sample of length 1000 and drew 10 nonoverlapping subsamples from it, each of length 100. We know the error rate of TE for data length 100 (converged using 10000 samples) from equations 19 and 20. Let us assume this rate to be constant for all the subsamples drawn and be denoted by $E_s$. Then, for an ensemble created with these subsamples, as in equation 16, the ensemble error rate ($E_e$) using binomial probability would be,

$$E_e = \sum_{i=9}^{10} C_i^{10} \times E_s^i \times (1 - E_s)^{10-i} \quad (21)$$

The ensemble error rates are calculated for all the $(m/\epsilon)$ ratios and are plotted (red dashed lines) along with the error rates obtained for data lengths 100 and 1000. Supplementary Figure 2 shows that an ensemble approach performs better than the parent sample (violet lines). An ensemble estimator outperforms the traditional TE, especially for samples with low $(m/\epsilon)$, which shows that our method increases the robustness of TE estimates against noise.

**Author Contributions:**

SG and AG conceived the idea and designed the problem. All the authors developed the solution algorithms. EE wrote the code and performed the analysis with the input from TC and VC. All the authors analyzed the results. AG, SG, and EE wrote the manuscript with input from all the authors.

**Competing Interests**

The authors declare no competing interests

**Acknowledgment**

The work presented here is financially supported by the Department of Science and Technology, Government of India, Project no: DST/ SJF/ E&ASA-01/2018-19; SB/SJF/2019-20/11. ARG's time was funded by US DOD's SERDP grant # RC 20-1183.

# Supplementary Information

## for

## Robust Causality and False Attribution in Data-Driven Earth Science Discoveries

**Elizabeth Eldhose[1], Tejasvi Chauhan[1], Vikram Chandel[2], Subimal Ghosh[1,2]\*, and Auroop R. Ganguly[3,4]**

[1]Department of Civil Engineering, Indian Institute of Technology Bombay, Mumbai, India

[2]Interdisciplinary Program in Climate Studies, Indian Institute of Technology Bombay, Mumbai, India.

[3]Sustainability and Data Sciences Laboratory, Department of Civil and Environmental Engineering, Northeastern University, Boston, MA, USA

[4]Pacific Northwest National Laboratory, Richland, WA, USA

**\* Corresponding Author: Email: subimal@civil.iitb.ac.in; subimal.ghosh@gmail.com**

**Supplementary Figures: 1 to 7**

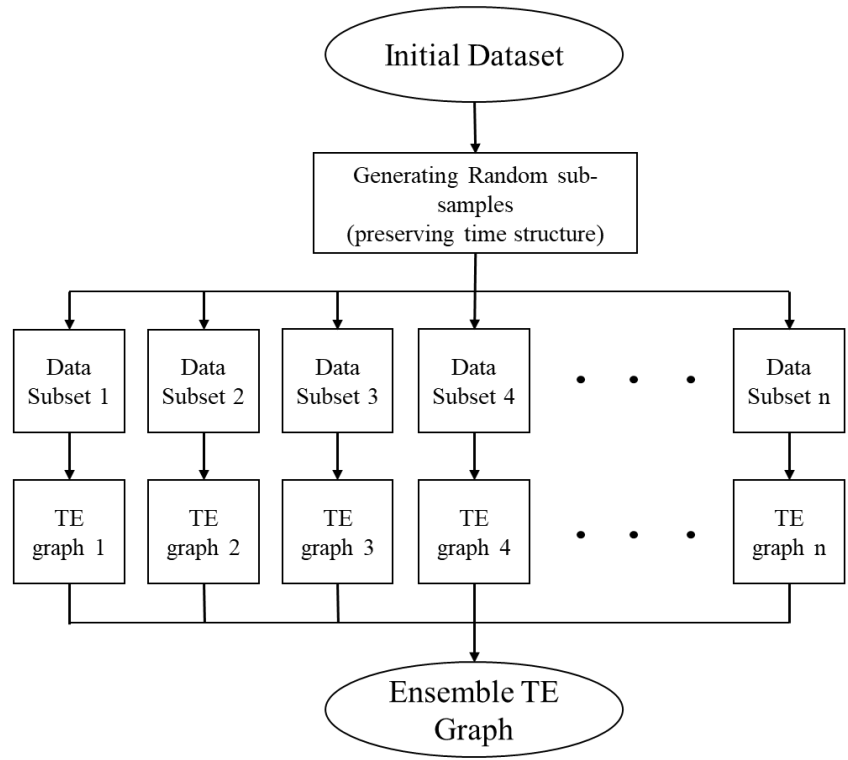

***Supplementary Figure 1:*** *Flow Chart for Ensemble Technique*

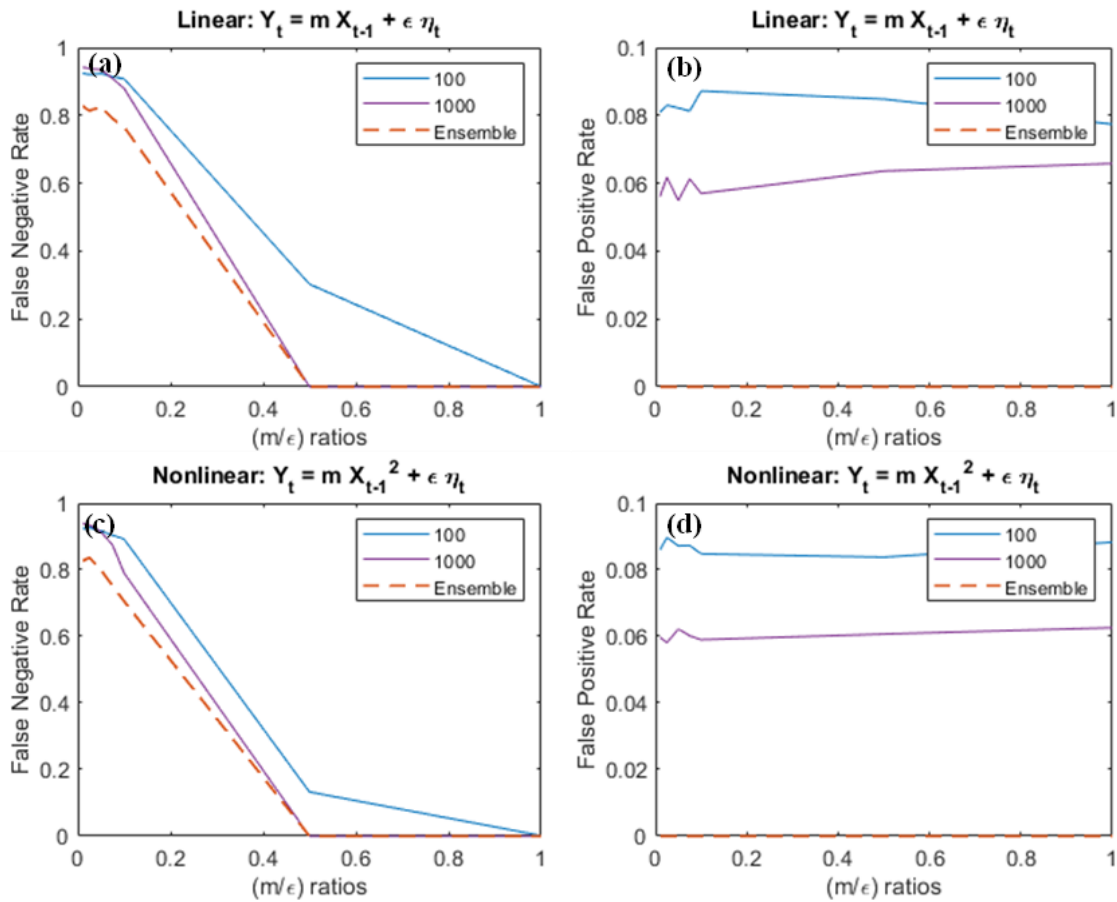

***Supplementary Figure 2:*** *False Negative and False Positive Rates plot as a function of varying $(m/\epsilon)$ ratios for data lengths 100, 1000, and an ensemble of 10 subsamples of length 100.*

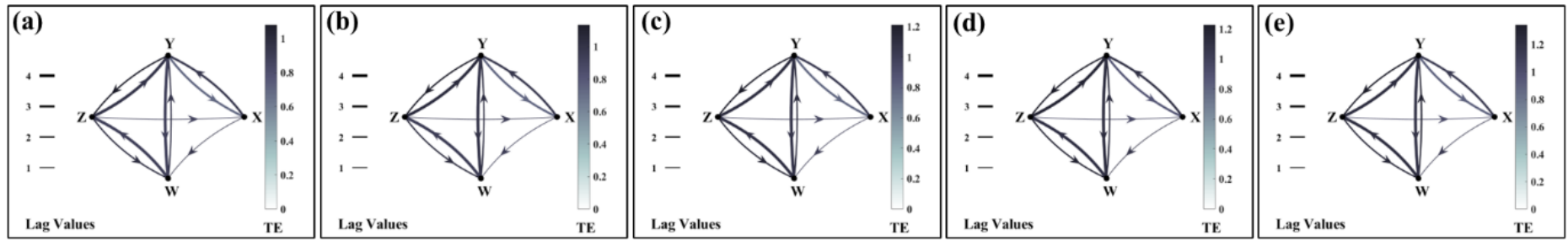

***Supplementary Figure 3:*** *Sensitivity of transfer entropy links over the number of bins used to estimate the probability distribution. Transfer Entropy graph for the System C with 900 data points (a) 18 bins (b) 19 bins (c) 20 bins (d) 21 bins (e) 22 bins. The number of bins obtained for the system was 20 bins using Scott's rule.*

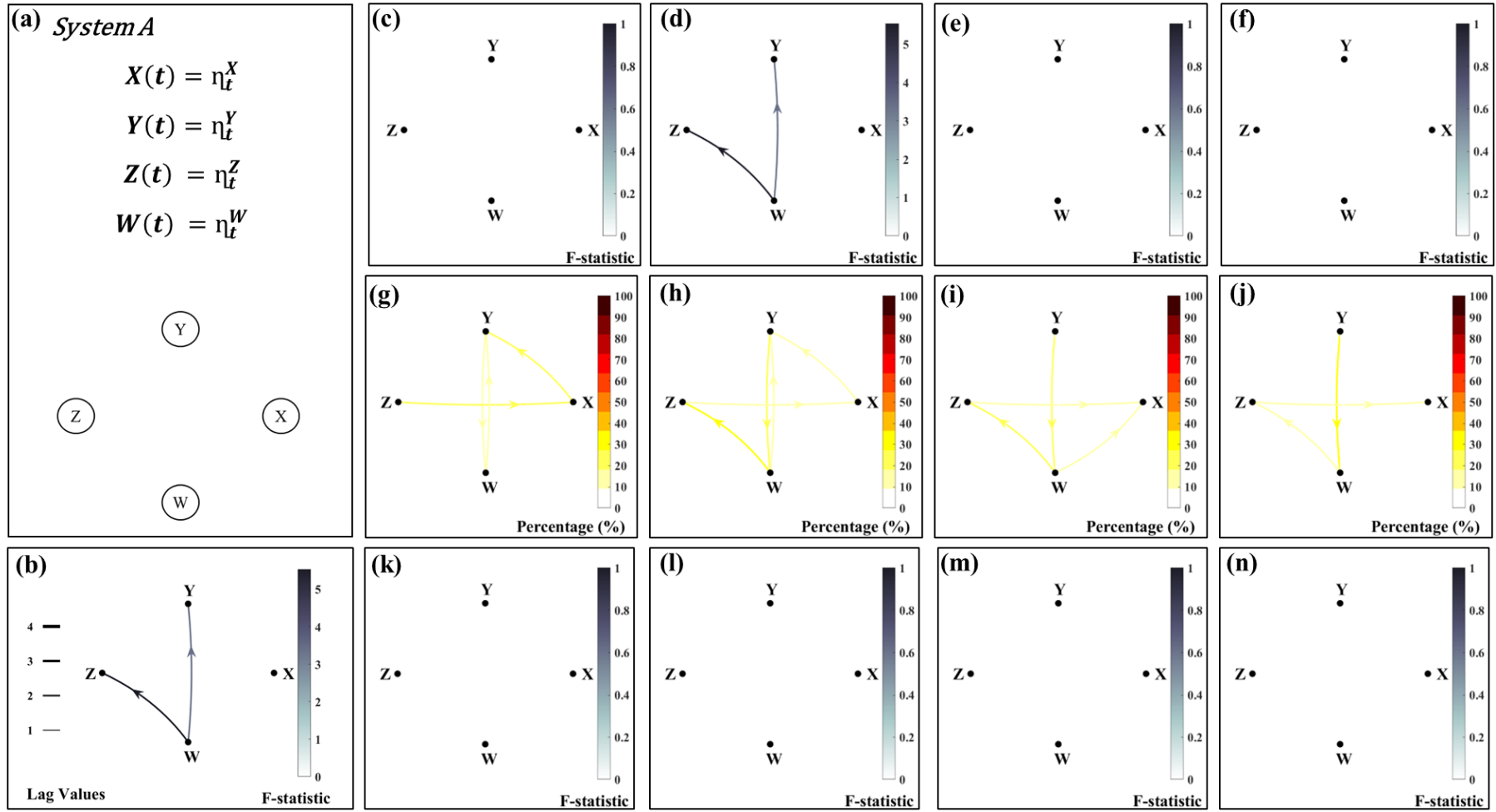

***Supplementary Figure 4:*** *(a) System A of four independent variables; X, Y, Z, and W. Each variable is a time series of normally distributed random numbers with 1000 data points. Plot (b) shows a directed graph with the bivariate Granger causality links up to lag 4, obtained from the sample of size 1000. Lag-wise Granger causality links are presented in (c) Lag 1, (d) Lag 2, (e) Lag 3, and (f) Lag 4. Figures (g) -(j) denote each link's appearance percentage at the four lags after applying granger causality to 100 subsamples of size 200 each. Correspondingly, the links that appeared above 90% times at four lags are shown in (k) - (n)*

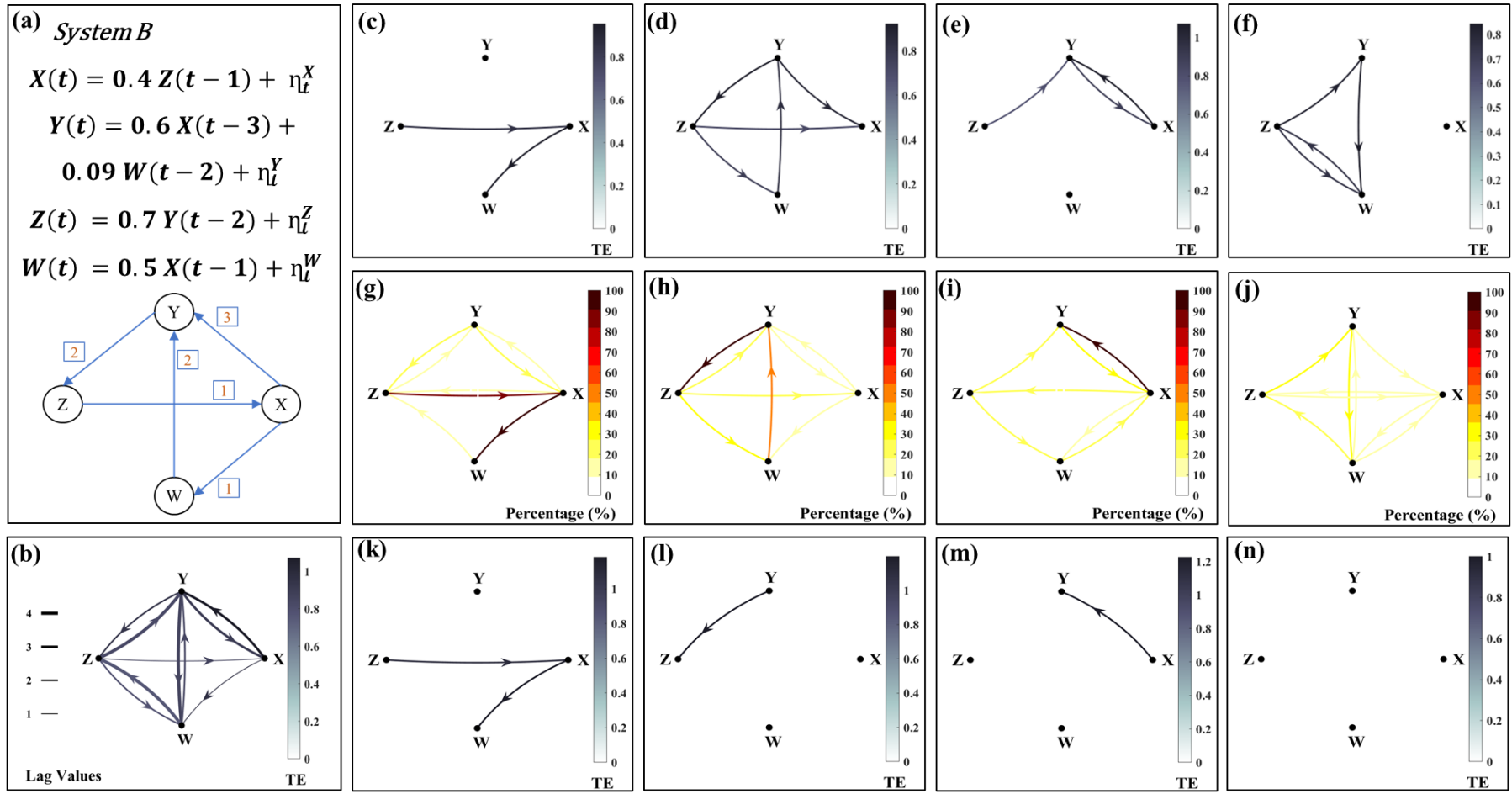

***Supplementary Figure 5:*** *(a) System B with four variables, X, Y, Z, and W, of 900 data points each. A sample of size 1000 is generated with the equations mentioned in (a). After discarding the first 100 values, the rest 900 values are used for further analysis. Plot (b) shows a directed graph with the significant transfer entropy links up to lag 4, obtained from the sample of size 900. Lag-wise transfer entropy links are presented in (c) Lag 1, (d) Lag 2 (e) Lag 3, (f) Lag 4. Figures (g) -(j) denote each link's appearance percentage at the four lags after applying transfer entropy to 100 subsamples of size 200 each. Correspondingly, the links that appeared above 90% times at four lags are shown in (k) - (n)*

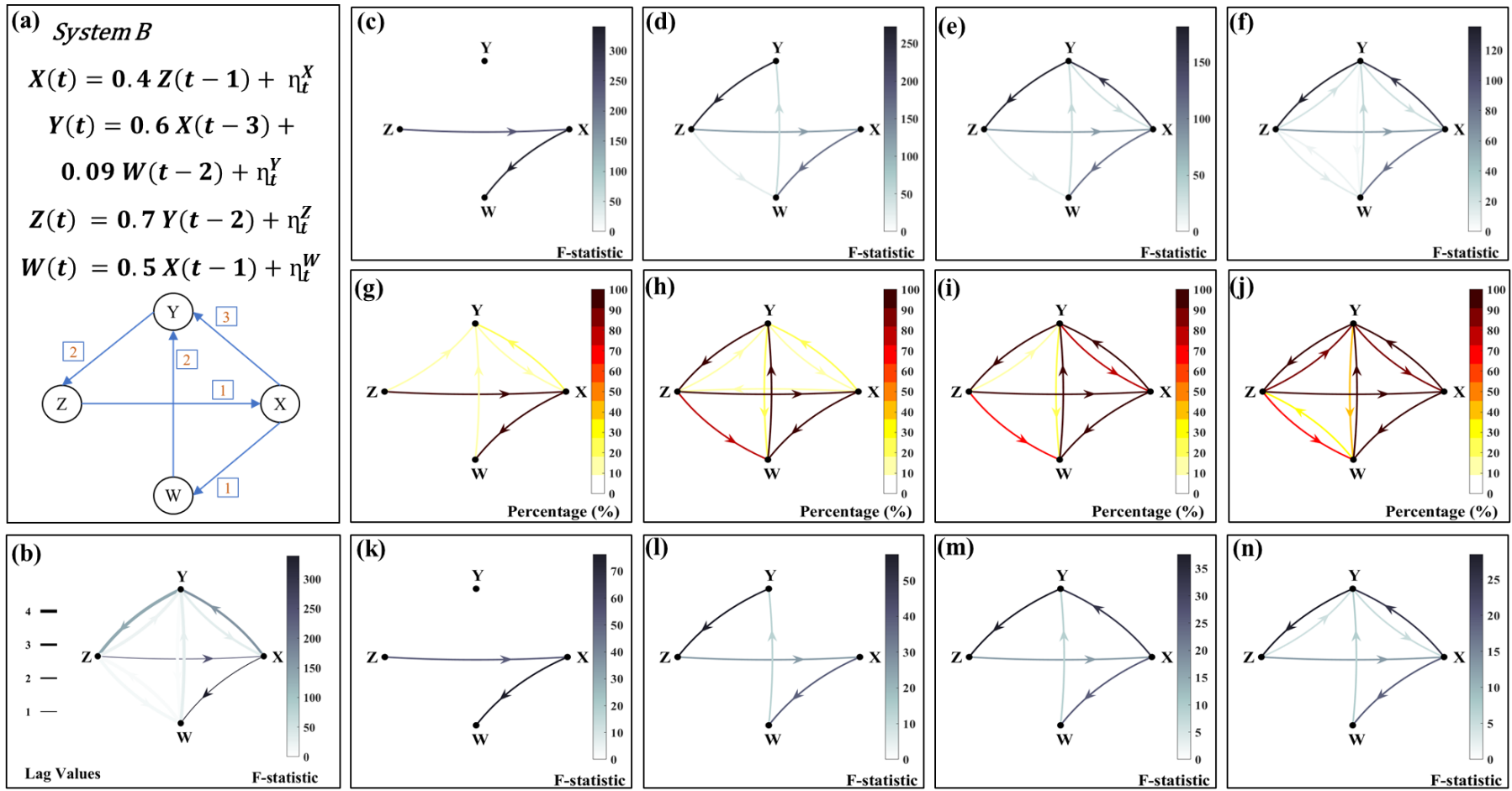

***Supplementary Figure 6:*** *(a) System B with four variables, X, Y, Z, and W, of 900 data points each. A sample of size 1000 is generated with the equations mentioned in (a). After discarding the first 100 values, the rest 900 values are used for further analysis. Plot (b) shows a directed graph with the bivariate Granger causality links up to lag 4, obtained from the sample of size 900. Lag-wise Granger causality links are presented in (c) Lag 1, (d) Lag 2 (e) Lag 3, and (f) Lag 4. Figures (g) -(j) denote each link's appearance percentage at the four lags after applying Granger causality to 100 subsamples of size 200 each. Correspondingly, the links that appeared above 90% times at four lags are shown in (k) - (n)*

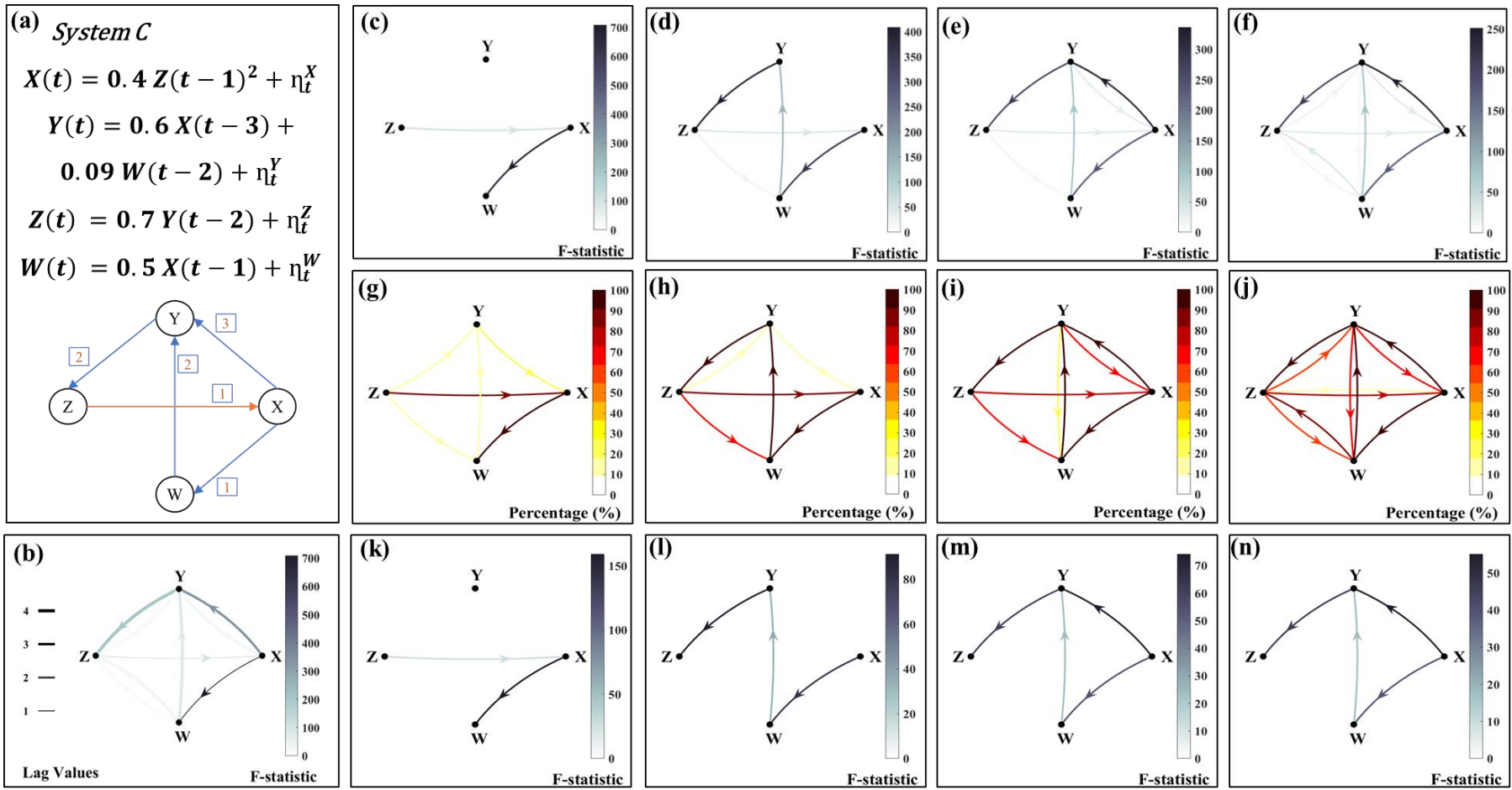

***Supplementary Figure 7:*** *(a) System C with four variables, X, Y, Z, and W, of 900 data points each. A sample of size 1000 is generated with the equations mentioned in (a). After discarding the first 100 values, the rest 900 values are used for further analysis. Plot (b) shows a directed graph with the bivariate Granger causality links up to lag 4, obtained from the sample of size 900. Lag-wise Granger causality links are presented in (c) Lag 1, (d) Lag 2 (e) Lag 3, and (f) Lag 4. Figures (g) -(j) denote each link's appearance percentage at the four lags after applying Granger causality to 100 subsamples of size 200 each. Correspondingly, the links that appeared above 90% times at four lags are shown in (k) - (n)*